\documentclass[aps,prl,twocolumn,showpacs,superscriptaddress]{revtex4-1}

\usepackage{graphicx}
\usepackage{dcolumn}
\usepackage{bm}
\usepackage{hyperref}
\usepackage[utf8]{inputenc}
\usepackage[T1]{fontenc}
\usepackage{lmodern}
\usepackage[english]{babel}
\usepackage{mathtools}
\hypersetup{colorlinks=true,citecolor={blue},linkcolor={blue},urlcolor={blue}}

\usepackage{amsmath}
\usepackage{empheq}
\usepackage{mathrsfs}
\usepackage{amssymb}
\usepackage{soul}
\usepackage{upgreek}

\begin{document}

\setlength{\parskip}{0mm plus0mm minus0mm} 


\title{Synchronization in optically-trapped polariton Stuart-Landau networks}

\author{S. L. Harrison}
\address{School of Physics and Astronomy, University of Southampton,  Southampton, SO17 1BJ, UK.}

\author{H. Sigurdsson}
\address{School of Physics and Astronomy, University of Southampton,  Southampton, SO17 1BJ, UK.}
\address{Skolkovo Institue of Science and Technology, Novaya St. 100, Skolkovo 143025, Russian Federation.}

\author{P. G. Lagoudakis}
\address{School of Physics and Astronomy, University of Southampton,  Southampton, SO17 1BJ, UK.}
\address{Skolkovo Institue of Science and Technology, Novaya St. 100, Skolkovo 143025, Russian Federation.}

\begin{abstract}
We demonstrate tunable dissipative interactions between optically trapped exciton-polariton condensates. We apply annular shaped nonresonant optical beams to both generate and confine each condensate to their respective traps, pinning their natural frequencies. Coupling between condensates is realized through the finite escape rate of coherent polaritons from the traps leading to robust phase locking with neighboring condensates. The coupling is controlled by adjusting the polariton propagation distance between neighbors. This permits us to map out regimes of both strong and weak dissipative coupling, with the former characterized by clear in-phase and anti-phase synchronization of the condensates. With robust single-energy occupation governed by dissipative coupling of optically-trapped polariton condensates, we present a system which offers a potential optical platform for the optimization of randomly connected $XY$ Hamiltonians.
\end{abstract}

\maketitle

{\it Introduction.} Studies on instabilities, synchronization, and pattern formation in systems of limit-cycle oscillators appear in many scientific disciplines such as hydrodynamics, biological ensembles, neuronal networks, nonlinear optics, Josephson junctions, and coupled Bose-Einstein condensates~\cite{Cross_RevModPhys1993, Acebron_RevModPhys2005, Matheny_Science2019}. In the regime of strong light-matter coupling, condensates of microcavity exciton-polaritons (herein polaritons) are found to follow similar oscillatory dynamics due to their nonlinear and dissipative physics~\cite{kavokin_microcavities_2011}. The condensation of polaritons~\cite{Deng_RevModPhys2010}, attributed to their bosonic nature and very light effective mass, has given rise to a powerful experimental platform to investigate nonlinear and out-of-equilibrium physics at the macroscopic quantum level and even at room-temperature~\cite{plumhof_room-temperature_2014}.

The dynamics between multiple coupled polariton condensates, denoted by a complex number $c_n$, are can be described using a discretized version of the driven-dissipative Gross-Pitaevskii equation (dGPE)~\cite{Lagoudakis_2010PRL, Stepnicki_PRA2013, Rayanov_2015PRL, kalinin2019polaritonic},
\begin{equation}\label{eqn:dc_dt0}
        i\frac{dc_n}{dt} = \Big[\Omega_n + \alpha |c_n|^2\Big]c_n+\sum_{\langle n m \rangle} J_{nm}c_m.
\end{equation}
where $\Omega_n=\omega_n + i (p_n - \gamma_n)$, $\alpha=g-iR$, and $J_{nm}=|J_{nm}|e^{i\beta_{nm}}$ denote the complex self-energy of each condensate, its non-linearity, and coupling to nearest neighbors respectively. Physically, $\gamma_n,p_n >0$ denote the condensate linear losses and gain respectively. For a single condensate the evolution of its density $|c|^2$ coincides with that of the Landau equation, describing the dynamics of disturbances in the laminar flow of fluids, $\partial_t |c|^2= k_1 |c|^2 + k_2 |c|^4$, where $k_{1,2}$ are real constants~\cite{landau1944problem, stuart_1960}. When connections are present, $J_{nm} \neq 0$, Eq.~\eqref{eqn:dc_dt0} can be regarded as a discretized form of the complex Ginzburg-Landau equation~\cite{Hakin_PRA1992} describing a system of coupled limit-cycle oscillators labeled as \emph{Stuart-Landau networks}. We note that the complex Ginzburg-Landau equation differs from the driven-dissipative Gross-Pitaevskii equation in its historical origin and intent~\cite{Aranson_RevModPhy2002}.

Interestingly, recent studies on optical networks of limit-cycle oscillators have found that there exists a regime with a strong attractor in phase space, where the relative difference between the arguments of the oscillators, $\theta_{nm} = \arg{(c_n^* c_m)}$, correlates with the ground state of the $XY$ Hamiltonian~\cite{Lagoudakis_NJP2017, berloff_realizing_2017, Kalinin_PRL2018}, and the Ising-Hamiltonian~\cite{Inagaki_NatPho2016, Inagaki_Science2016}. However, in order for an optical system to work in this ``minimal spin energy'' regime, the natural frequencies of the oscillators $\omega_n$ need to be resonant with each other, and $J_{nm}$ should be imaginary valued to ensure that dissipative coupling between oscillators fixes a definite phase relationship~\cite{kalinin_global_2018}. From a practical viewpoint, the relative phases in a desynchronized network of limit-cycle oscillators would average to zero over time. Thus, the phase information cannot be extracted in any setup relying on time-average measurements. It therefore becomes paramount, in order to successfully extract the phase configurations $\theta_{nm}$, that one possesses enough control over the networks parameters for it to remain synchronized such that phase readout is possible.

In this paper, we experimentally demonstrate and analyze an optical system of limit-cycle oscillators with tunable couplings $J_{nm}$ and fixed global natural frequencies $\omega_n = \omega$ using optically confined exciton-polariton condensates. We demonstrate clear regimes of synchronization between two condensates and map these regimes to the weights of the $XY$ Hamiltonian. We corroborate our findings by numerically solving both the continuous and discretized version of the driven-dissipative Gross-Pitaevskii equation, and benchmark the dGPE's performance in finding the $XY$ ground state against system uncertainties.

The optically trapped condensates are formed by exciting a semiconductor microcavity with a non-resonant laser pump profile shaped into rings~\cite{askitopoulos_polariton_2013, askitopoulos_robust_2015}. The ring-shaped pumps start building up trapped polaritons which at a critical power form a condensate in the minimum of the pump potential. Because of their non-equilibrium nature, polaritons can diffuse away from their pumping spots, transforming their potential energy into kinetic energy. Such a flow of coherent polaritons~\cite{Schmutzler_PRB2015, Su_Science2018, toepfer_neuromorphic_2019}, with tunable cavity in-plane momentum, then leads to interference and robust phase locking between spatially separated condensates~\cite{Wouters_PRB_Synchronized_2008, Baas_PRL_Synchronized_2008, Eastham_PRB_Mode_2008, Christmann_NJP2014, ohadi_nontrivial_2016}. 

The resulting phase locking can be detected by observation of interference fringes in the cavity real-space and/or reciprocal-space photoluminescence, which has been achieved today over more than one hundred microns~\cite{toepfer_neuromorphic_2019}. More importantly, optically trapped polariton condensates show coherence time which exceeds the cavity lifetime by 3 orders of magnitude~\cite{Akitopoulos_giant_2019}, increasing the scalability of the system to phase lock far beyond that of the optically pumped regime. Moreover, by scaling to a condensate network interspersed with optically imprinted variable-height potential barriers~\cite{alyatkin_optical_2019}, we propose a robust platform on which to imprint nearly arbitrary weights belonging to an $XY$ Hamiltonian into the polariton system for heuristic optical ground state searching through nonlinear transients.

{\it Experiment.} We experimentally realise the trapped polariton condensates using a strain compensated $2\lambda$ GaAs planar microcavity sample containing 3 pairs of InGaAs quantum wells sandwiched between another InGaAs quantum well pair, as described in \cite{cilibrizzi_polariton_2014}. The sample is held in a cold finger cryostat at $\sim$4 K and is non-resonantly pumped at a cavity-detuning of around -5meV by right-circularly polarised light from a continuous wave (CW) Ti:Sapphire laser, blue-detuned in energy to a minimum above the reflectivity stopband ($\lambda = 780$ nm). To avoid heating the sample, we form a quasi-CW beam with the use of an acousto-optic modulator, at a 10 KHz repetition rate and 5\% duty cycle, to modulate the amplitude of the beam periodically. The annular shape of the beam is achieved using a spatial light modulator (SLM) displaying a phase-modulating hologram (see Supplementary Material for method of hologram generation). The beam  is focused on the surface of the sample using two lenses and a high numerical aperture objective (NA$ = 0.4$). The photoluminescence emission is collected through the same objective and an $808$ nm long pass filter is used to cut out the excitation beam. The beam is also spectrally resolved with a 1800 grooves/mm grating in a 750 mm spectrometer, centered at 857 nm.

\begin{figure}[t!]
    \centering
    \includegraphics[width=1\columnwidth]{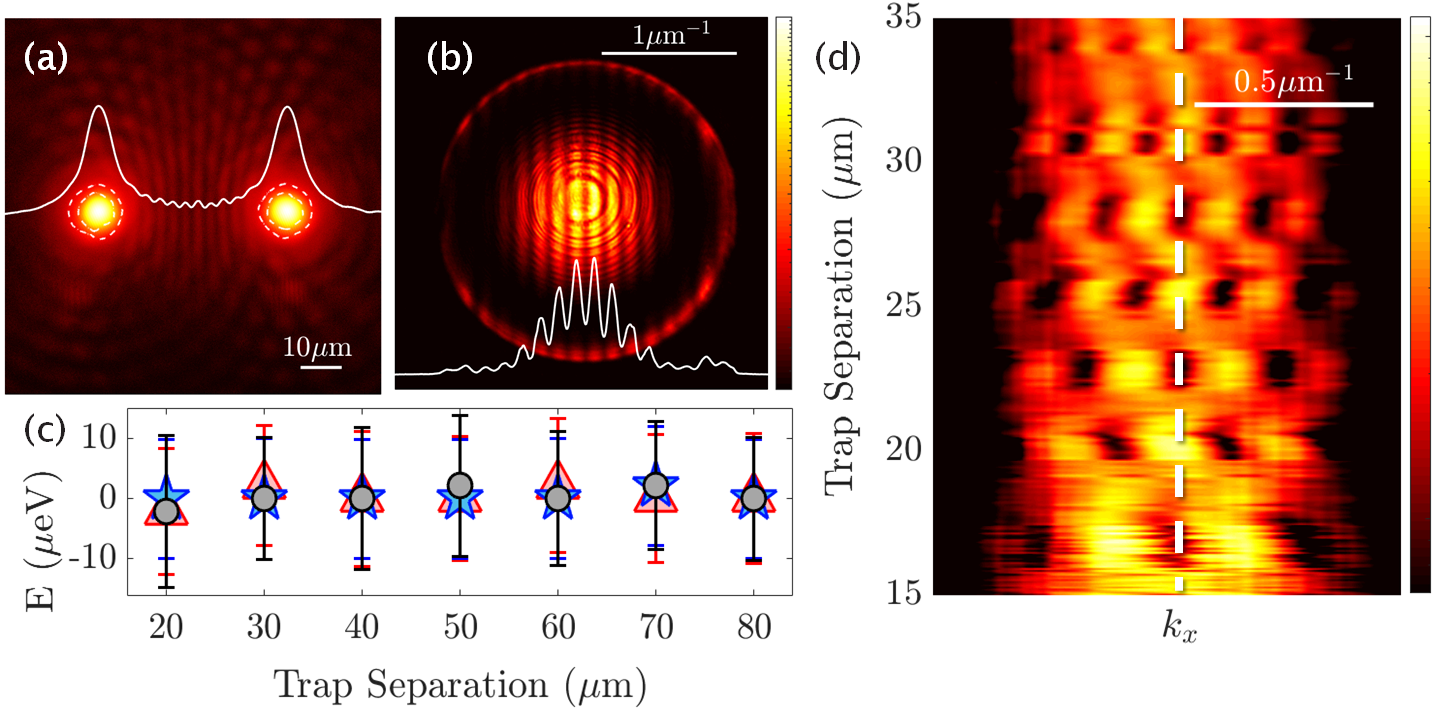}
    \caption{(a) Real- and (b) reciprocal-space condensate photoluminescence for two annular traps at $d = 50$ $\upmu$m separation. Integrated horizontal line profiles are shown by white lines. (c) Energies of the condensates for varying separation.  Red triangles and blue pentagrams correspond to uncoupled left and right condensate respectively. Black circles correspond to coupled condensates. Error bars show FWHM of the energy linewidth. Zero energy corresponds to 0.796 meV above the bottom of the lower polariton dispersion. (d) Integrated horizontal line-profiles from reciprocal-space photoluminescence (i.e., at $k_y = 0$) for varying trap separation. Dashed white marks $k_x = 0$.}
    \label{fig:rr_kk_manyfringes}
\end{figure}

{\it Results and Discussion.} Above threshold power, polaritons condense into a phase coherent trap ground state at the pump center. In Fig.~\ref{fig:rr_kk_manyfringes}(a, b) we show the real-space and reciprocal-space condensate photoluminescence respectively for two phase locked condensates as evidenced by the clear formation of interference fringes. The radial outflow of coherent polaritons from their pumping spots corresponds to the faint outer ring seen in reciprocal-space whereas the brighter central region corresponds to polaritons localized in the traps. In Fig.~\ref{fig:rr_kk_manyfringes}(d) we plot the integrated horizontal line-profile in reciprocal space, taken over separation distances (i.e., the real space distance between the ring centers) from 15 $\upmu$m to 35 $\upmu$m for rings $d_\text{trap} = 9.4$ $\upmu$m in diameter. The interference fringes indicate that the condensates are phase locked, where a bright or dark central fringe shows even (in-phase) and odd (anti-phase) parity respectively.

The observed phase locking means that the coupling between condensates cannot be negligible and should therefore result in normal mode splitting, where new energies of the system are shifted away from the bare energies of the uncoupled system. Surprisingly for the trapped condensates studied here, we observe that the energy of the coupled system stays within the linewidth of the polaritons [see Fig.~\ref{fig:rr_kk_manyfringes}(c)]. This observation is made more clear by considering the interacting condensates in the linear regime as a zero-detuned two-level system with states $|c_1\rangle$ and $|c_2\rangle$ and energy $\omega_{1,2} = \omega$. Coupling between the states is realized with an operator of the form $\hat{\mathcal{J}} = J e^{i\beta} \hat{\sigma}_1$, where $J>0$, and the Hamiltonian becomes,
\begin{equation} \label{eq.H}
\mathcal{H} = \omega \hat{\sigma}_0 + \hat{\mathcal{J}} = 
\begin{pmatrix}
\omega & Je^{i\beta} \\
Je^{i\beta} & \omega
\end{pmatrix}.
\end{equation}
Here, $\hat{\sigma}_n$ are the Pauli matrices. The resulting even and odd parity eigenmodes of Eq.~\eqref{eq.H} (corresponding to in-phase and anti-phase locking), written $\Psi = (1,\pm1)^T/\sqrt{2}$, will have eigenfrequencies $\omega_{A,B} = \omega \pm 2 J e^{i \beta}$. It is clear if $\beta = \pm \pi/2$ then both modes are degenerate in real frequency but are split by $\Delta = i2J$ in imaginary frequency (i.e., their linewidths are different). During condensation, a state of definite parity will form corresponding to the eigenstate with a larger imaginary part in its energy. Physically, it corresponds to increased scattering from the reservoir of uncondensed polaritons, and consequently becomes populated during the transient process of condensation. Therefore, even though the real energy splitting is within the linewidth of the system [Fig.~\ref{fig:rr_kk_manyfringes}(c)], the impact of the dissipative splitting is not negligible as evidenced by the clear regions of interference fringes, indicating condensation into a definite parity [Fig.~\ref{fig:rr_kk_manyfringes}(d)].

As can be seen in Fig.~\ref{fig:rr_kk_manyfringes}(d), regions of clear interference fringes appear periodically as a function of separation distance with intermediate transition regions of no clear parity. This periodic behavior stems from the fact that away from the pumped rings the polariton flow is dictated by solutions of the time-independent cylindrical wave equation (i.e., the Helmholtz equation) which are given by the Hankel functions~\cite{Wouters_PRB2008}. This results in $J_{nm}$ spiraling in the complex plane to smaller values with increasing polariton outflow momentum and distance between traps~\cite{Lagoudakis_NJP2017, berloff_realizing_2017}. When the coupling is dominantly imaginary ($\beta = \pm\pi/2$) then fringes appear clearly due to deterministic condensation into the highest gain mode. When the coupling is dominantly real ($\beta = 0,\pi$) then both parity modes are degenerate in gain and stochastically condense, where by phase-locking can occur with either even or odd parity for each realisation of the system, rather than both at once. With a camera exposure time of $\sim$1~ms, multiple realisations are measured with each experimental shot, with half randomly forming in even parity states, and the other in odd. This results in the blurring seen in Fig.~\ref{fig:rr_kk_manyfringes}(d) as both parity states are realised, smearing out the interference fringes in the shot-to-shot averaged measurements of the experiment.
\begin{figure}[b!]
    \centering
    \includegraphics[width=1\columnwidth]{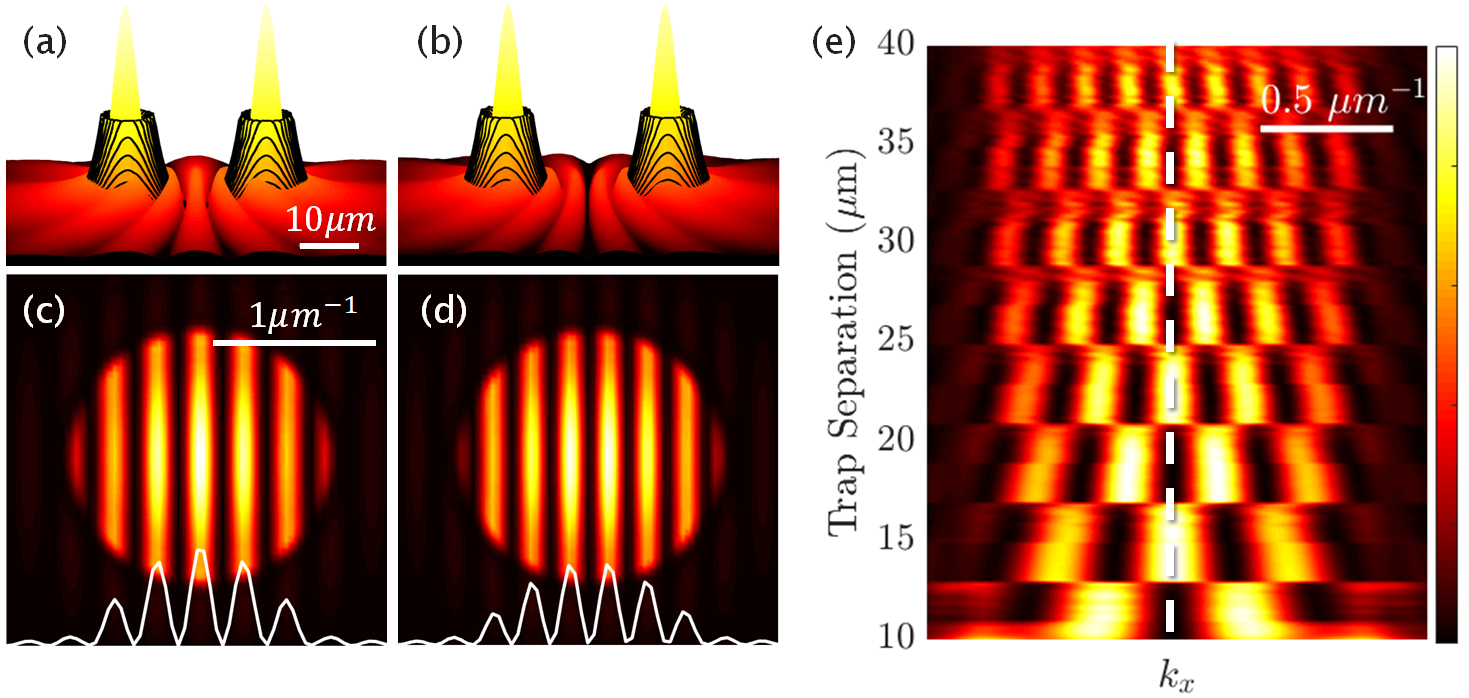}
    \caption{Simulations obtained through the full numerical solution of the 2DGPE in real-space (a,b) and reciprocal-space (c,d) density profiles of two interfering steady state polariton condensates. In-phase locking is observed at separation distance of 23.2 $\upmu$m (a,c), and anti-phase locking at 26.6 $\upmu$m (b,d). All plots are shown on the same normalized color scale. Black-contoured surface in (a,b) illustrated the trap potentials formed by the incident laser profiles. White lines in (c,d) show the horizontal line-profiles of the reciprocal-space density. (e) Horizontal reciprocal-space density line-profiles shown for varying trap separation distances.}
    \label{fig:simulation_2}
\end{figure}

The above findings are corroborated by numerical simulations using the two-dimensional driven-dissipative Gross-Pitaevskii equation (2DGPE) [see Supplementary Material]. In agreement with experiment, the energy of the simulated condensate wavefunction maintains, on average, a value around $\sim 0.782$ meV above the bottom of the lower polariton branch, and parity switching is seen in both real-space and reciprocal-space condensate profiles as the separation distance varies (see Fig.~\ref{fig:simulation_2}(a-d) for two different steady state examples). In Fig.~\ref{fig:simulation_2}(e), the horizontal line-profile in reciprocal space is plot for two ring traps of $d_\text{trap} = 7$ $\upmu$m diameter with separation distances going from 10 to 40 $\upmu$m. This figure is built up by averaging the line profile over 20 simulation realisations, each starting from a different random background noise. In agreement with experiment, we see smeared-out interference fringes in the transition regions when the dissipative coupling is weak (i.e., coupling becomes $\beta \approx 0,\pi$). We note that we do not apply a time-dependent stochastic treatment of the 2DGPE. Consequently, the simulation [Fig.~\ref{fig:simulation_2}(e)] shows a much sharper transition from one parity to next as opposed to the extended blurred regions seen in experiment.
\begin{figure}[t!]
    \centering
    \includegraphics[width=1\columnwidth]{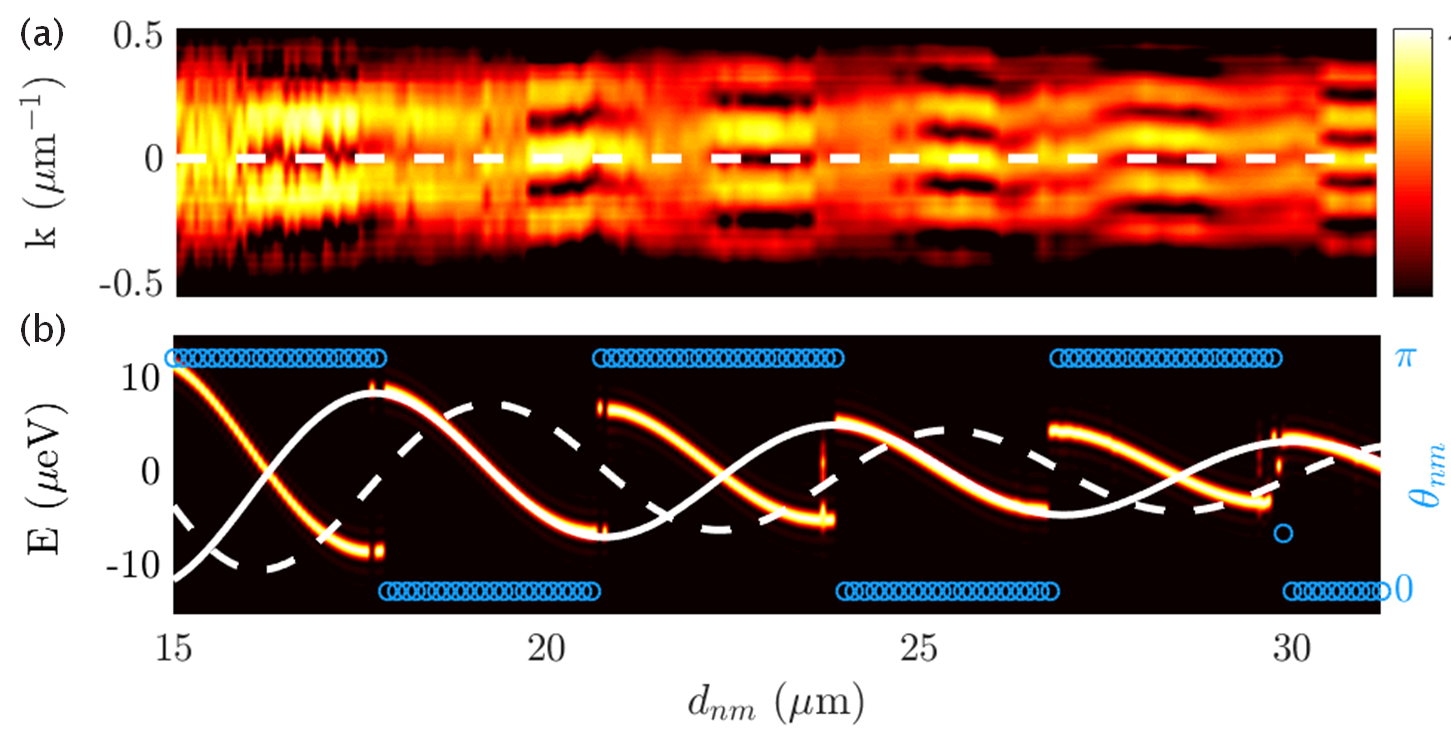}
    \caption{(a) Experimental reciprocal-space photoluminescence as shown in  Fig~\ref{fig:rr_kk_manyfringes}(d), with dashed white line to show $k = 0$ $\upmu$m$^{-1}$. (b) Colormap shows the spectral intensity (real energy) from simulation of two condensates as a function of distance $d$ using Eq.~\eqref{eqn:dc_dt0} and with coupling as defined by Eq.~\eqref{eqn:Hankel}. Blue circles show the steady state relative phase $\theta_{12}$ from same simulation. Solid and dashed white lines correspond to the real and imaginary parts of Eq.~\eqref{eqn:Hankel} respectively. Parameters: $k_c = 1.04$ $\upmu$m$^{-1}$, $d_\text{trap} = 9.4$ $\upmu$m, $\mu = 0.28$ meV ps$^2$ $\upmu$m$^{-2}$, $\omega_n = 0$ and $\gamma_n^{-1} = 5.5$ ps, $p_n/\gamma_n = 1.65$, $R = 0.005$ ps$^{-1}$, and $g/R = 0.02$.}
    \label{fig:theory_sep}
\end{figure}

We now describe the observations using Eq.~\eqref{eqn:dc_dt0} which can be derived by adiabatically eliminating the dynamics of the exciton reservoirs feeding the condensates~\cite{Keeling_PRL2008} and applying a tight binding method for the localized dissipative condensates~\cite{Stepnicki_PRA2013, kalinin2019polaritonic}. The coupling is taken proportional to the Hankel function (see Supplementary Figures 2,3),
\begin{equation}
J_{nm} = |J_{nm}|e^{i\beta_{nm}} = J_0 e^{i \phi} H_0^{(1)}[k(d_{nm} - d_\text{trap})].
\label{eqn:Hankel}
\end{equation}
Here $J_0$ is the magnitude of the coupling strength, $\phi$ is a phase adjustment parameter to match experiment, $k = k_c + i \mu \gamma_n/2\hbar k_c$ where $k_c$ is the outflow polariton momentum~\cite{Wouters_PRB2008}, $\mu$ is the polariton mass, $d_{nm}$ is the separation distance between condensate $n$ and $m$, and $d_\text{trap}$ is the trap diameter.

The condensation of two interacting polariton condensates is then simulated using Eq.~\eqref{eqn:dc_dt0} and the resulting spectral intensity (real energy) for $p_n>\gamma_n$ is plotted in Fig.~\ref{fig:theory_sep}(b) as a function of separation distances varying from 15 $\upmu$m to 32 $\upmu$m. The blue circles denote the relative phase between condensates $\theta_{12} = \arg{(c_1^*c_2)}$ showing step-function regions of in-phase and anti-phase locking. In Fig.~\ref{fig:theory_sep}(a) we show a section from Fig.~\ref{fig:rr_kk_manyfringes}(d) for comparison. The spectrum shows discontinuous jumps where the imaginary (dissipative) part of the coupling $J_{nm}$ (dashed white curve) changes sign. This corresponds to the lowest threshold condensate mode switching parities. The solid white curve denotes the real part of $J_{nm}$. The results show that the system of two coupled condensates follows robustly the highest gain mode dictated by the imaginary part of $J_{nm}$.

By additionally modulating the phase of the coupling ($\beta_{nm}$) through the use of optically generated potential-barriers~\cite{alyatkin_optical_2019}, the couplings $J_{nm}$ between adjacent condensates can be programmed to have nearly arbitrary values of magnitude $|J_{nm}|$, with phases chosen as $\beta_{nm} \approx \pm \pi/2$. This then allows the design of a synchronized random network of dissipative coupled limit-cycle oscillators for simulation of the $XY$ Hamiltonian~\cite{Lagoudakis_NJP2017, berloff_realizing_2017, Kalinin_PRL2018}. Applying Eq.~\eqref{eqn:dc_dt0}, the principle idea is starting with $p_n- \gamma_n$ negative enough that  $c_n = 0$ is the only stable solution of the network. Physically, this scenario corresponds to condensates being pumped below threshold. By adiabatically increasing $p_n$ (slowly raising the pump power), this fixed point eventually becomes unstable and the system undergoes a nonlinear transient process (Hopf bifurcation) to a ``condensed'' steady state $|c_n| > 0$ whose phase configuration correlates with that of the $XY$ ground state.

\begin{figure}[t!]
    \centering
    \includegraphics[width=1\columnwidth]{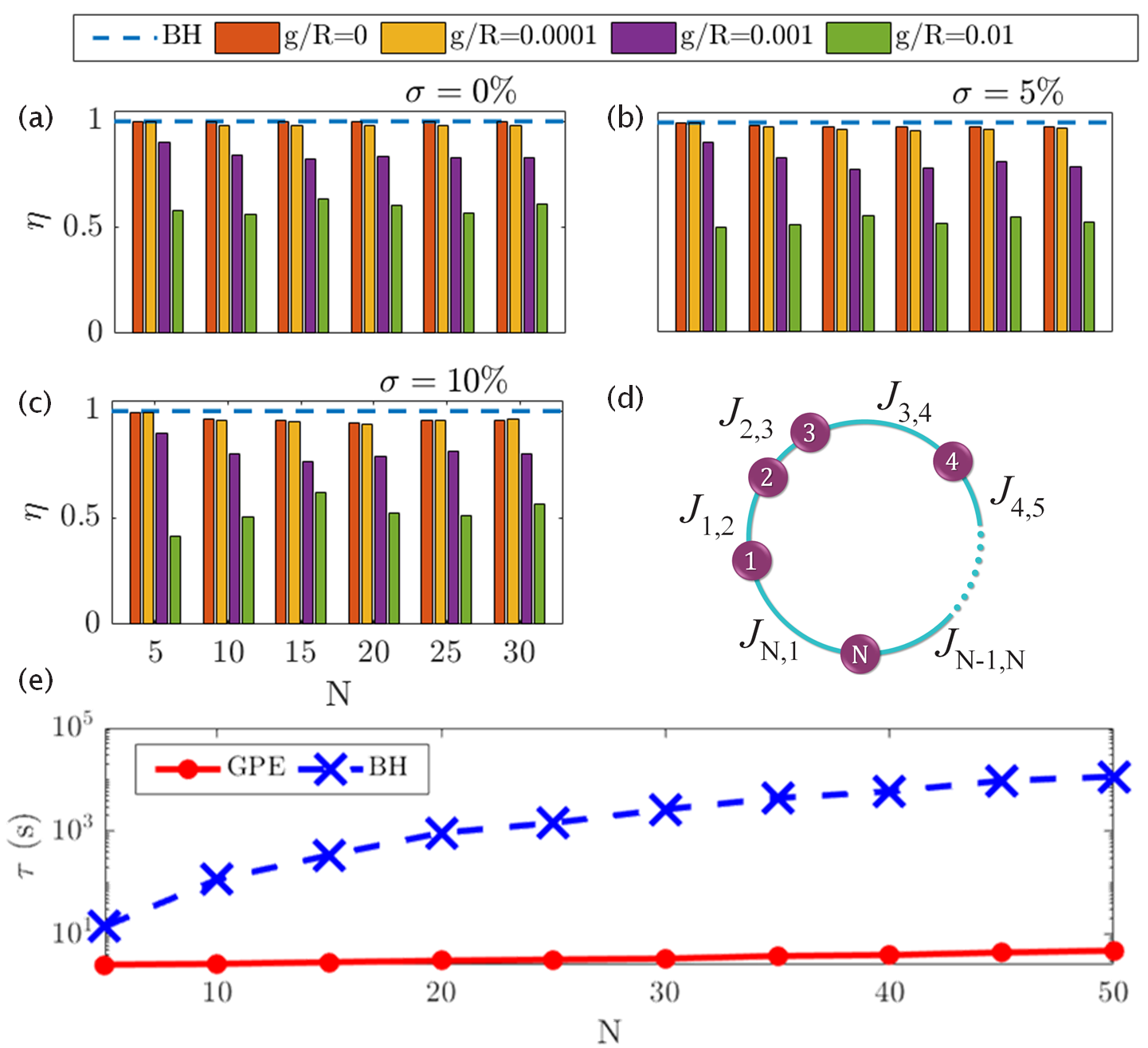}
    \caption{(a-c) Average ratio of minimised $XY$ Hamiltonian energy between the dGPE and BH method, $\eta = E_\text{GPE}/E_\text{BH}$, for 20 random realizations of couplings $J_{nm}$. Performance is shown for different network sizes $N$, values of $g/R$, and $\sigma$. (d) shows schematic of the randomly-connected continuous-chain lattice tested with dGPE and BH. (e) Elapsed computation time taken, $\tau$, to find a local minimum energy phase configuration using the dGPE and BH methods.}
    \label{fig:simulation_3}
\end{figure}

In a comparable method to~\cite{kalinin_global_2018,Kalinin_NJP2018}, we verify the performance of Eq.~\eqref{eqn:dc_dt0} and test it against a global classical optimizer, the Basin Hopping (BH) method~\cite{Wales_BasinHopping_1997}, at finding the $XY$ Hamiltonian ground state of a randomly connected closed chain [see Fig.~\ref{fig:simulation_3}(d)], $\mathcal{H}_{XY} = -\sum_{nm} \text{Im}{(J_{nm})} \cos{(\theta_{nm})}$. Since no cavity system is ideal, the robustness of the dGPE is additionally investigated by deviating $\beta_{nm}$ from the ideal values of $\pm \pi/2$. We also investigate the effects of the ratio of the two nonlinearities $g/R$ where $g$ is responsible for shifting the real energy of each condensate [see Eq.~\eqref{eqn:dc_dt0}]. Results on a fully connected random etwork of condensates is given in Supplementary Figure 4. Illustrative phase configuration for a network of 10 randomly connected spins after minimising the $XY$ Hamiltonian via the dGPE and Basin Hopping method is shown in Supplementary Figure 5.
    
For each set of undirected couplings $J_{nm}$, we define the energy found by the BH method and the dGPE as $E_\text{BH}$ and $E_\text{GPE}$ respectively. In Fig.~\ref{fig:simulation_3}(a-d) we plot a histogram of $\eta = E_\text{GPE}/E_\text{BH}$ averaged over 20 different random coupling configurations for different numbers of condensates $N$ in the network. Figure~\ref{fig:simulation_3}(a) shows the results in an ideal case where the phases of the couplings are randomly chosen either $\beta_{nm} = \pm \pi/2$. In Fig.~\ref{fig:simulation_3}(b,c) we plot the performance with deviation in the couplings defined as $J_{nm} = |J_{nm}| (e^{\pm i \pi/2} \pm \sigma )$ where the plus-minus signs are randomly chosen separately. The deviation $\sigma>0$ then corresponds to the dissipative coupling obtaining a small real part which can desynchronize the network of condensates.

With $g/R= 0.0001$, the performance does not drop below $\eta = 0.96$ and we see that whilst $g \leq R$, $\eta$ remains above 0.78 and does not vary significantly as $\sigma$ changes. Increasing $g$ beyond this point considerably reduces the accuracy of the dGPE. The computation time for each method is also shown in Fig.~\ref{fig:simulation_3}(e) over different sized systems, where each minimization method is implemented for a single coupling configuration on a single core of the same Intel(R) Xeon(R) W3520 @ 2.67GHz CPU. The computational time taken by the dGPE increases by just 2.5 seconds as the system size is scaled by a factor of 10, while time taken by the classical BH method scales more than three orders of magnitude.

{\it Conclusions.} We have demonstrated robust synchronization between optically-trapped polariton condensates which is attributed to a dissipative coupling mechanism arising from the condensates mutual interference. The coupled condensate system does not show a measurable normal mode splitting due to the linewidth of the polaritons, yet at the same time displays ability to synchronize at separation distances where dissipative coupling is dominant. The single-frequency operation of the system is critical in order to read out the relative phase information between interacting condensates in time-average measurements. It therefore offers a way to implement the recently proposed gain-dissipative Stuart-Landau networks for ultrafast simulation of randomly connected spin Hamiltonians in the optical regime~\cite{Lagoudakis_NJP2017,berloff_realizing_2017,Kalinin_PRL2018}.\\

{\it Acknowledgements.} The authors acknowledge technical support from Mr. Julian T\"{o}pfer and Dr. Ioannis Chatzopoulos. PGL acknowledges useful discussions with Prof. Nikolay A. Gippius for recognizing the importance of controlling the natural frequencies of ballistically expanding coupled condensates. The authors  acknowledge the support of the UK’s Engineering and Physical Sciences Research Council (grant EP/M025330/1 on Hybrid Polaritonics), the use of the IRIDIS High Performance Computing Facility,  and associated support services at the University of Southampton.

 \newcommand{\noop}[1]{}

\newpage
\setcounter{figure}{0}
\setcounter{equation}{0}
\renewcommand{\theequation}{S\arabic{equation}}
\renewcommand{\thefigure}{S\arabic{figure}}
\renewcommand{\thesection}{S\arabic{section}}

\begin{center}
\Large{\textbf{Supplemental Material}}
\end{center}

\section*{Numerical Spatiotemporal Simulations}
The dynamics of polariton condensates can be modelled via the mean field theory approach where the condensate order parameter $\Psi(\mathbf{r},t)$ is described by a 2D semiclassical wave equation often referred as the generalised Gross-Pitaevskii equation coupled with an excitonic reservoir which feeds non-condensed particles to the condensate~\cite{Wouters_Excitation_2007}. The reservoir is divided into two parts: an active reservoir $n_{\mathrm{A}}(\mathbf{r},t)$ belonging to excitons which experience bosonic stimulated scattering into the condensate, and an inactive reservoir $n_{\mathrm{I}}(\mathbf{r},t)$ which sustains the active reservoir~\cite{Lagoudakis_Coherent_2010,Lagoudakis_Probing_2011}.
\begin{align}
i  \frac{\partial \Psi}{\partial t} & = \left[ -\frac{\hbar \nabla^2}{2m} + \frac{G}{2} (n_{\mathrm{A}}+n_{\mathrm{I}}) +  \frac{U}{2} |\Psi|^2 + \frac{i }{2} \left( \xi n_{\mathrm{A}} - \gamma \right) \right] \Psi,  \label{eq.GPE} \\ 
\frac{ \partial n_{\mathrm{A}}}{\partial t} & = - \left( \Gamma_{\mathrm{A}} + \xi |\Psi|^2 \right) n_{\mathrm{A}} + W n_{\mathrm{I}}, \label{eq.ResA} \\
\frac{ \partial n_{\mathrm{I}}}{\partial t} & = -  \left(\Gamma_{\mathrm{I}} + W \right) n_{\mathrm{I}} + P(\mathbf{r}). \label{eq.ResI}
\end{align}
Here, $m$ is the effective mass of a polariton in the lower dispersion branch, $U$ is the interaction strength of two polaritons in the condensate, $G$ is the polariton-reservoir interaction strength, $\xi$ is the rate of stimulated scattering of polaritons into the condensate from the active reservoir, $\gamma$ is the polariton decay rate, $\Gamma_{{\mathrm{A,I}}}$ is the decay rate of active and inactive reservoir excitons respectively, $W$ is the conversion rate between inactive and active reservoir excitons, and $P(\mathbf{r})$ is the non-resonant CW pump profile.

We perform numerical integration of Eqs.~\eqref{eq.GPE}, \eqref{eq.ResA} and~\eqref{eq.ResI} in time using a linear multistep method in time and spectral methods in space. The polariton mass and lifetime are based on the sample properties: $m = 0.28$ meV ps$^2$ $\upmu$m$^{-2}$ and $\gamma = \frac{1}{5.5}$ ps$^{-1}$. We choose values of interaction strengths typical of InGaAs based systems: $\hbar U = 7$ $\upmu$eV  $\upmu$m$^2$, $G = 10 U$. The non-radiative recombination rate of inactive reservoir excitons  is taken to be much smaller than the condensate decay rate ($\Gamma_{\mathrm{I}} = 0.01 \gamma$), whereas the active reservoir is taken comparable to the condensate decay rate $\Gamma_{\mathrm{A}} = \gamma$ due to fast thermalisation to the exciton background~\cite{Wouters_Spatial_2008}. The final two parameters are then found by fitting to experimental results where we use the values $\hbar \xi = 99$ $\upmu$eV $\upmu$m$^{2}$, and $W = 0.035$ ps$^{-1}$.

\section*{Annular Pump Profiles}
The pump profiles consist of two annular traps, each written as $P(r) = P_0 e^{-(r -r_0)^2/2\sigma^2}$ where $P_0$ denotes the pump power, $r$ sweeps radially from the centre of each pump, $r_0$ marks the trap radius and $\sigma$ corresponds to a $2\;\mathrm{\upmu m}$ experimentally diffraction-limited full width at half maximum of each annulus. 

\begin{figure}[t!]
    \centering
    \includegraphics[width=0.6\columnwidth]{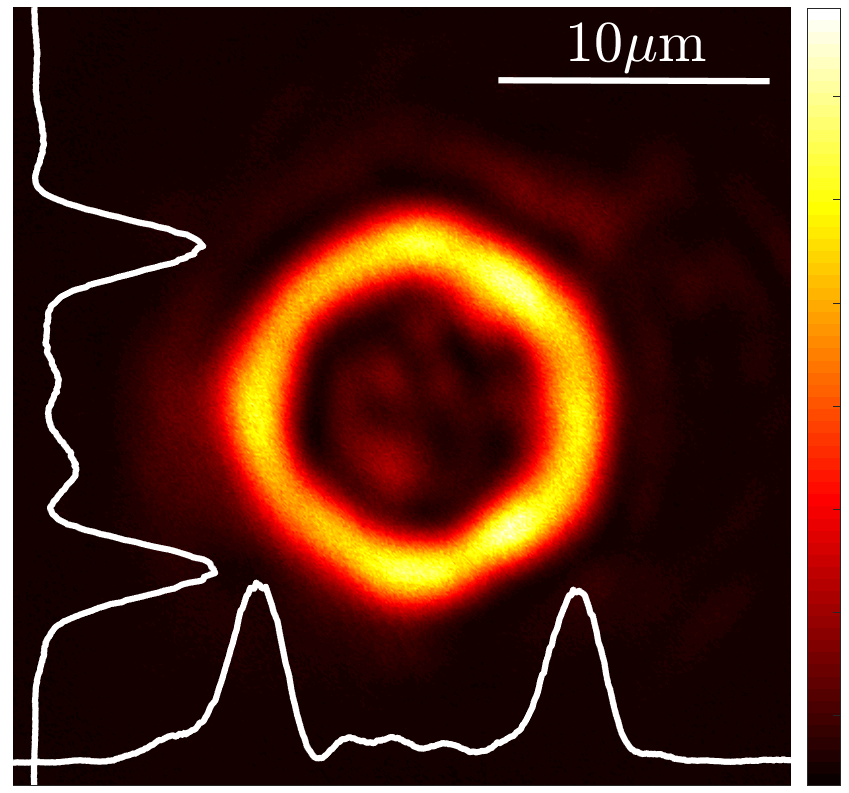}
    \caption{Real-space laser photo-luminescence of 12 $\upmu$m diameter pump, including horizontal and vertical line profiles through the centre of the pump profile, shown by solid white lines.}
    \label{fig:suppl1}
\end{figure}

Experimentally, the annular shape of the beam is achieved using a spatial light modulator (SLM) displaying a phase-modulating hologram. The hologram is created using the mixed-region amplitude-freedom (MRAF) algorithm~\cite{Pasienski_2007}, and adjusted to balance the condensate intensities~\cite{Nogrette_Single-Atom_2014}. The laser photo-luminescence profile of a pump used to trap a single polariton condensate in its ground state is shown in~\ref{fig:suppl1}.

\section*{Hankel function polariton outflow}
\begin{figure}[t!]
        \centering
        \includegraphics[width=1\columnwidth]{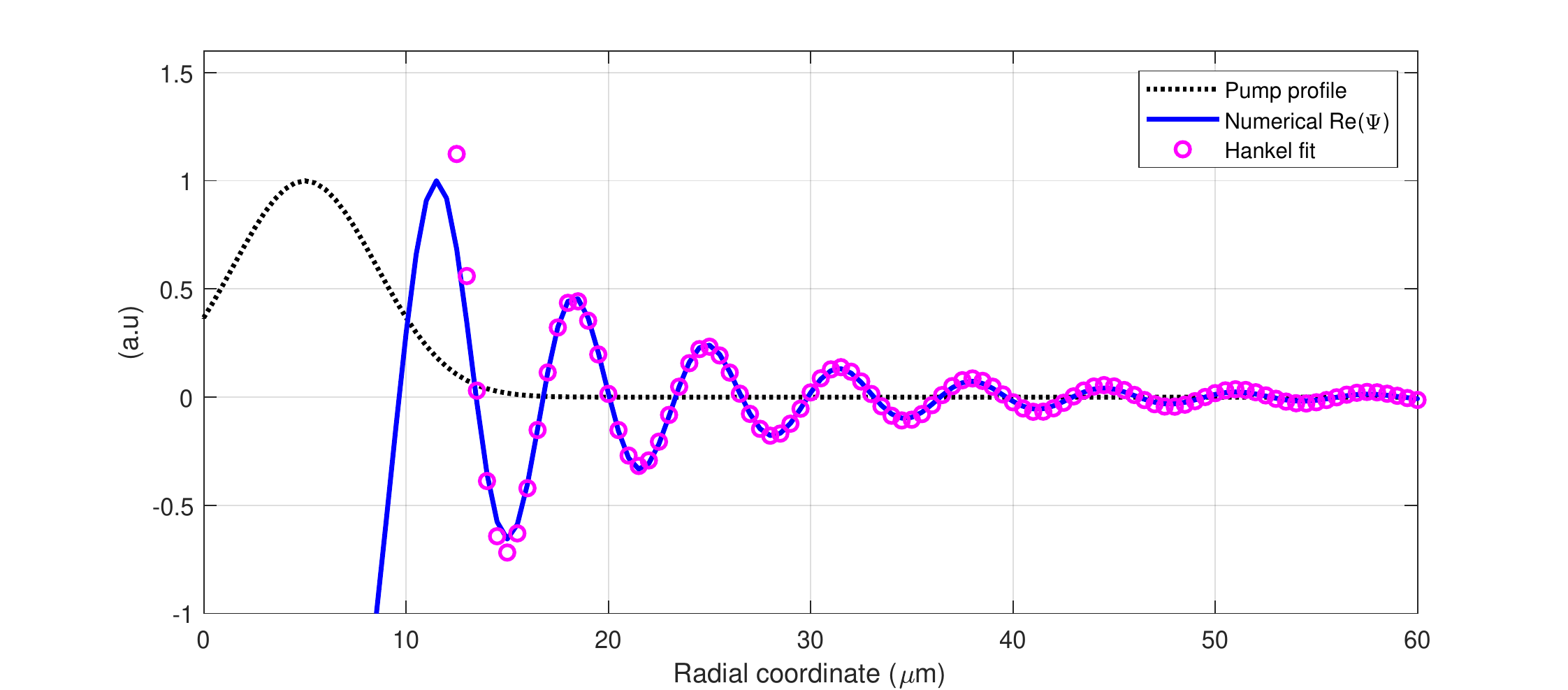}
        \caption{Cross-section of the nonresonant pump profile (black dotted line), real part of the numerically obtained condensate wavefunction from 2DGPE simulations, and a fitted zeroth-order Hankel function of the first kind $\Psi(\mathbf{r}) = A e^{i \phi} H_0^{(1)}[k(r-r_0)]$, where $A$ and $\phi$ are real valued fitting parameters. Radial coordinate corresponds to $r$. Here $k = k_c + i \kappa$ where $k_c = 0.96$ $\upmu$m$^{-1}$ and $\kappa = m\gamma /2 \hbar k_c$. The parameter $r_0=11$ $\upmu$m is adjusted to the point where particles have escaped the trap.}
        \label{fit}
    \end{figure}

In~\ref{fit} we fit a zeroth-order Hankel function of the first kind (magenta circles) to a steady state solution of Eqs.~\eqref{eq.GPE}-\eqref{eq.ResI} (solid blue curve) for a annular shaped pump geometry (black dotted line). The results show that outside of the pump spot the steady state condensate assumes the solution of the Helmholtz equation as expected. 

\begin{figure}[t!]
        \centering
        \includegraphics[width=1\columnwidth]{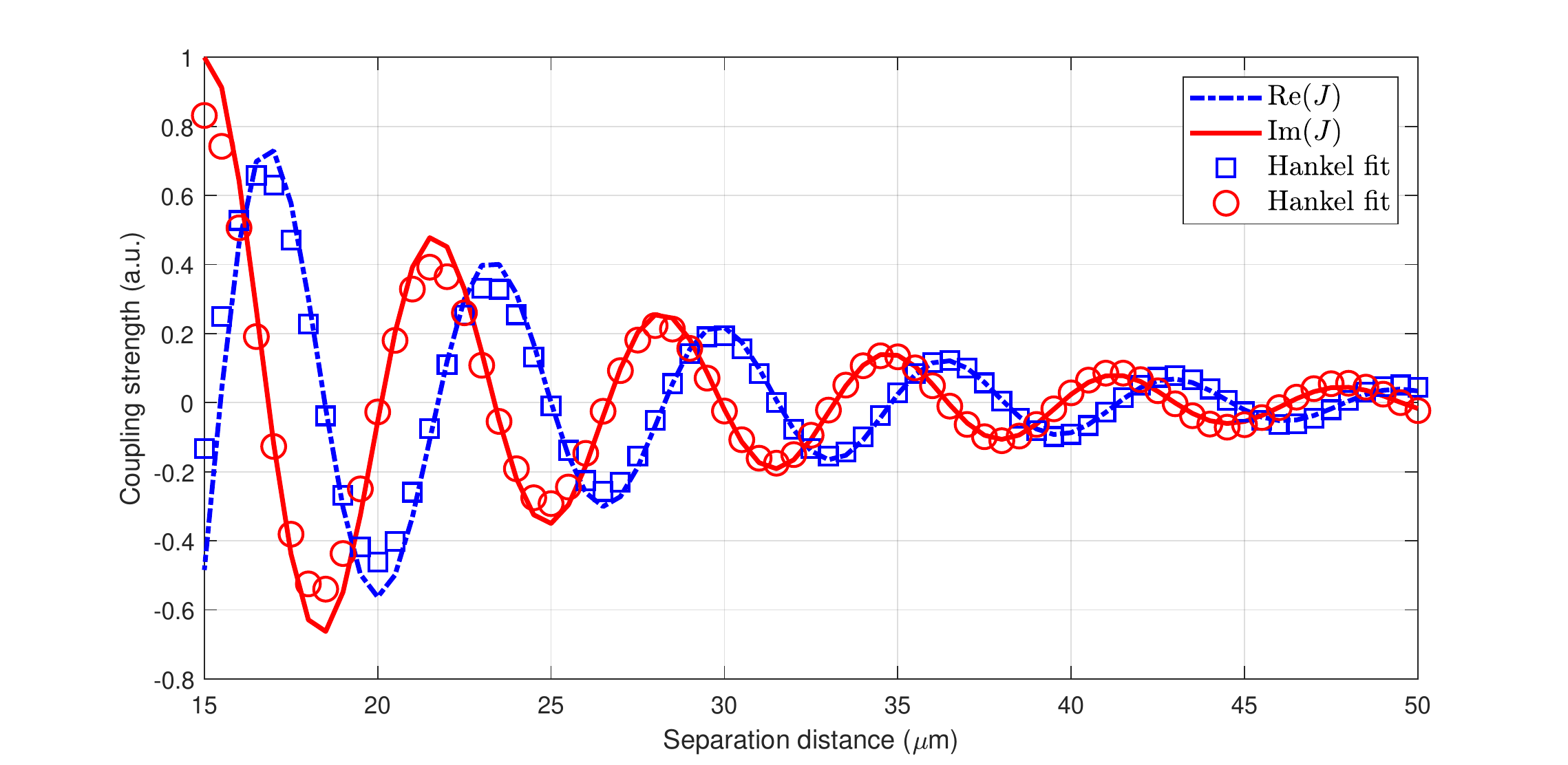}
        \caption{Real and imaginary part from Eq.~\eqref{int} in reply letter for  wavefunctions separated by a distance $d = |\mathbf{d}|$. As the magnitude of the distance is decreased the value of the integral expectedly decreases as the overlap diminishes. The circle and square markers are a fit to the data using a Hankel function $J = J_0 e^{i \phi} H_0^{(1)}[k(d-d_\text{trap})]$. Here $k = k_c + i \kappa$ where $k_c = 0.96$ $\upmu$m$^{-1}$, $\kappa = m\gamma /2 \hbar k_c$, and $d_\text{trap} = 10$ $\upmu$m is the diameter of the trap used in this simulation.}
        \label{J}
    \end{figure}

Moreover, we verify the validity of approximating the coupling with a Hankel function [see Eq.~(3) in main text] by calculating the overlap integral between two condensates using the 2DGPE steady state solution of a single condensate,
\begin{equation} \label{int}
    J = \int \psi^*(\mathbf{r}-\mathbf{d}) V(\mathbf{r}) \psi(\mathbf{r}) d\mathbf{r}.
\end{equation}
Here $\psi(\mathbf{r})$ is the numerically obtained steady state condensate wavefunction for a single pump system by solving the 2DGPE, $V(\mathbf{r})$ is its corresponding optical trap, and $\mathbf{d}$ is the separation between two such neighboring wavefunctions. The results of the integration as a function of separation distance $|\mathbf{d}|$ are shown in~\ref{J} where we fit a zeroth-order Hankel function of the first kind (red circles and blue squares) to the values obtained from Eq.~\eqref{int}. The results show that the precise details of the pump shape are not necessary and that a qualitative analytical form to the coupling between condensates can be obtained by considering their wavefunction shape outside the pumped potential.
    
\section*{Densely Connected Polariton Graph}
\begin{figure*}[t!]
    \centering
    \includegraphics[width=2\columnwidth]{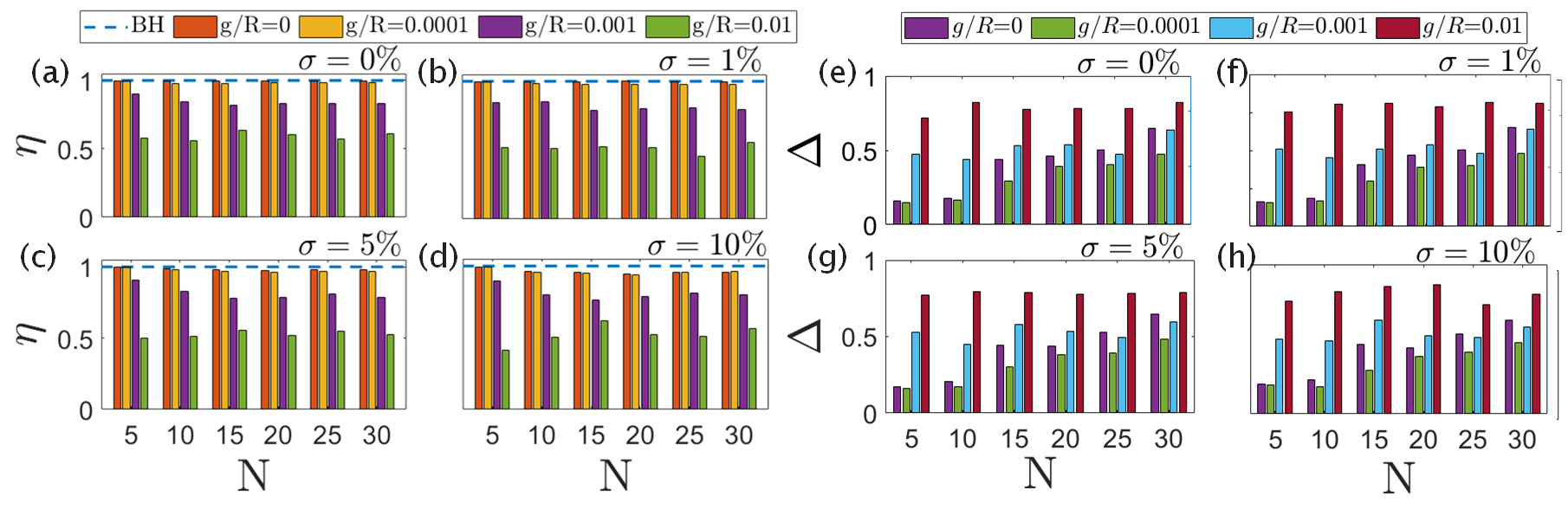}
    \caption{(a-d) measure of $\eta = E_{dGPE}/E_{BH}$ for a range of polariton-polariton interaction strengths $g/R$, and fraction of non-dissipative coupling strength $\sigma$ for a range of $N$ densely and randomly coupled spin. (e-h) Average standard deviation $\Delta$ ($M=20$) between the minimized spins using the dGPE and BH methods, again for a range of $g/R$ and $\sigma$ for different $N$.}
    \label{fig:suppl_eng_std_alltoall}
\end{figure*}

\begin{figure*}[t!]
    \centering
    \includegraphics[width=1.5\columnwidth]{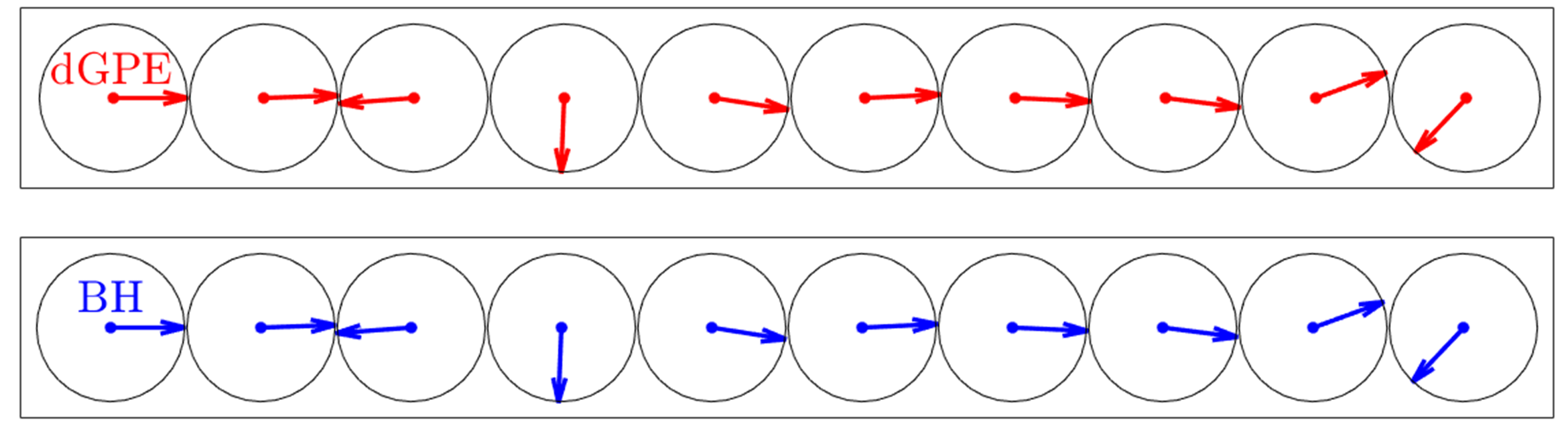}
    \caption{Relative phases of 10 randomly and densely connected spins minimising the $XY$ Hamiltonian achieved by (top) the dGPE and (bottom) the classical Basin Hopping method. In the minimisation shown, $g/R = 0.0001$ and $\sigma = 0\%$.}
    \label{fig:suppl2}
\end{figure*}

In addition to the closed sparsely-connected chain studied in the main text, we also compare the robustness of the dGPE to BH for a densely and randomly connected polariton graph of $N$ condensates [\ref{fig:suppl_eng_std_alltoall}(a-d)]. We plot $\eta = E_{dGPE}/E_{BH}$ for a range of realistic and unrealistic polariton-polariton interactions strengths and include a small percentage, $\pm\sigma$ of non-dissipative coupling to each value of $J_{nm}$, where $\pm$ is chosen randomly for each spin site. The minimisation of this all-to-all connected toy model, though unrealistic, shows that the dGPE is able to minimise any lattice configurations. An example of the minimised phases of the dGPE and BH is shown in~\ref{fig:suppl2}.

The average standard deviation between the dGPE and the BH is written:
\begin{equation}
 \Delta = \frac{1}{M} \sum_{m=1}^M \text{min} \sqrt{ \frac{1}{2N} (\mathbf{x}_\text{dGPE} - \mathbf{x}_\text{BH}^{(\pm)})^\dagger (\mathbf{x}_\text{dGPE} - \mathbf{x}_\text{BH}^{(\pm)})  },
\end{equation}
where $\mathbf{x}_\text{dGPE} = \{ e^{i \theta_n} \}_{n=1}^N$ and $\mathbf{x}_\text{BH}^{(\pm)} = \{ e^{\pm i \theta_n'} \}_{n=1}^N$ are the complex state vectors coming from each method with angles (phases) $\theta_n$ and $\theta_n'$ respectively. The global gauge is fixed by rotating the state vectors such that $\theta_1,\theta_1' = 0$ in each method. The min operation is added since the $XY$ Hamiltonian is invariant by an overall sign factor, i.e., $\mathcal{H}_{XY} = -\sum_{nm} \text{Im}{(J_{nm})} \cos{[\pm (\theta_{n} - \theta_m)]}$. The integer $M$ denotes the number of coupling realizations in the ensemble average (number of different networks tested).

In~\ref{fig:suppl_eng_std_alltoall}(e-h), we plot a histogram for $M=20$ realizations of random couplings $J_{nm}$ for $N$ condensates in the network.~\ref{fig:suppl_eng_std_alltoall}(e) shows the results in an ideal case where the phases of the couplings are randomly chosen either $\beta_{nm} = \pm \pi/2$. In~\ref{fig:suppl_eng_std_alltoall}(f-h) we plot the performance with deviation in the couplings defined as $J_{nm} = |J_{nm}| (e^{\pm i \pi/2} \pm \sigma )$ where the plus-minus signs are randomly chosen separately. The deviation $\sigma>0$ then corresponds to the dissipative coupling obtaining a small real part which can desynchronize the network of condensates. The results show that difference between the BH and the dGPE states increases when both $g$ and $\sigma$ increase. This then corresponds to the system becoming desynchronized.

\def\bibsection{\section*{Supplementary References}} 

\end{document}